\documentclass[conference]{IEEEtran}
\IEEEoverridecommandlockouts

\newcommand{\tcell}[1]{\begin{tabular}[x]{@{}l@{}}#1\end{tabular}}

\usepackage{cite}

\ifCLASSINFOpdf
\else
\fi

\usepackage{amsmath,amssymb}
\usepackage{color}
\usepackage[caption=false,font=footnotesize]{subfig}
\usepackage{lipsum,booktabs,multirow}
\usepackage{url}
\usepackage{amsmath}
\usepackage{graphicx}
\usepackage{enumitem}

\makeatletter
\def\input@path{{./tables/}}
\makeatother
\graphicspath{{./figures/}}

\newcommand{\Oo}{\mathcal{O}}
\newcommand{\bs}{\, \backslash \,}
\newcommand{\nnz}{\mathrm{nnz}}

\newcommand{\vbus}{v_{\mathrm{bus}}}
\newcommand{\sbus}{s_{\mathrm{bus}}}
\newcommand{\ibus}{i_{\mathrm{bus}}}
\newcommand{\Ybus}{Y_{\mathrm{bus}}}
\newcommand{\Zbus}{Z_{\mathrm{bus}}}
\newcommand{\CI}{\mathrm{CI}_{95\%}}

\newcommand{\vsrc}{v_{\mathrm{src}}}

\begin{document}

\title{Demonstrating Almost Linear Time Complexity of Bus Admittance Matrix-Based Distribution Network Power Flow: An Empirical Approach
\thanks{M. Deakin was supported by the Royal Academy of Engineering under the Research Fellowship programme, and by a CDICE Research Fellowship.}
}

\author{\IEEEauthorblockN{Matthew Deakin}
\IEEEauthorblockA{School of Engineering\\
Newcastle University\\
Newcastle-upon-Tyne, UK \\
matthew.deakin@newcastle.ac.uk}
\and
\IEEEauthorblockN{Davis Montenegro}
\IEEEauthorblockA{Power Delivery and Utilization\\
EPRI\\
Knoxville, TN, USA \\
dmontenegro@epri.com}
}

\maketitle

\begin{abstract}
The bus admittance matrix is central to many power system simulation algorithms, but the link between problem size and computation time (i.e., the time complexity) using modern sparse solvers is not fully understood. It has recently been suggested that some popular algorithms used in distribution system power flow analysis have cubic complexity, based on properties of dense matrix numerical algorithms; a tighter theoretical estimate of complexity using sparse solvers is not immediately forthcoming due to these solvers' problem-dependent behaviour. To address this, the time complexity of admittance matrix-based distribution power flow is considered empirically across a library of 75 networks, ranging in size from 50 to 300,000 nodes. Results across four admittance matrix-based methods suggest complexity coefficient values between 1.04 and 1.12, indicating complexity that is instead almost linear. The proposed empirical approach is suggested as a convenient and practical way of benchmarking the scalability of power flow algorithms.
\end{abstract}


\IEEEpeerreviewmaketitle

\section{Introduction}

\IEEEPARstart{U}{nbalanced} distribution system power flow is a core computational technique that is used by utilities to design, plan and operate distribution systems. Widespread uptake of distributed energy resources has led to growing interest in larger network models \cite{dugan2011open,enwl2015low,deakin2021hybrid,mateo2020building,deakin2023partitioned,koirala2020non} that capture new system interdependencies. The large scale of these problems has motivated the development in new methods that can exploit problem structure \cite{montenegro2022simplified,deakin2021analysis,kardovs2022beltistos}, enabling simulations over longer time periods with increased temporal resolution.

Methods building on the sparse bus admittance matrix $ \Ybus $ are an important class of algorithms used for power flow simulation. The $ \Ybus $ matrix links the vector of nodal voltage phasors $ \vbus $ and current phasors $ \ibus $ as
\begin{equation}\label{e:ybus_defn}
\Ybus \vbus = \ibus\,.
\end{equation}
Whilst exact algorithms vary, methods that use the $ \Ybus $ as the basis for distribution network power flow include fixed point-based Implicit $ \Zbus $ methods for non-linear power flow \cite{chen1991distribution,bazrafshan2017comprehensive}, Newton-Raphson based approaches using a power flow Jacobian \cite{garcia2000three,penido2013new}, as well as more recent power flow linearizations such as Fixed Point and First Order Taylor methods \cite{bernstein2018load}. For example, the popular OpenDSS tool (regularly used to benchmark new algorithms \cite{bernstein2018load,carreno2022log,bazrafshan2017comprehensive}) uses a $ \Ybus $-based Fixed Point method \cite{dugan2011open}.

In this work, we explore the computational complexity of distribution network power flow analysis as a `big-O' time complexity problem in the number of nodes $ n $. If it is assumed that the time to solve a solve power flow problem $ t $ is a polynomial of degree $ \alpha $ in the number of nodes $ n $, then
\begin{equation}\label{e:complexity_defn}
t \propto n^{\alpha }\,.
\end{equation}
Sparse network formulations of power flow problems have traditionally been favored as compared to dense formulations due to lower computational requirements. However, despite its central role in power engineering, power flow computational complexity has seldom been studied directly \cite{safdarian2022power}.

As a result in this gap, estimates of the value of $ \alpha $ vary widely. For example, it has recently been proposed that methods such as the fixed-point method of OpenDSS can have cubic complexity, i.e., $ \alpha=3 $ \cite{carreno2022log}. This contrasts with legacy texts that state that the complexity is linear, i.e., that $ \alpha=1 $ \cite{arrillaga1990computer}[Ch. 2.9]. In \cite{alvarado1976computational}, the authors statistically estimate the value of $ \alpha $ to be between 1.2 and 1.4, although only considering systems with up to 500 buses (and using pre-1980s hardware). To the authors' knowledge, no prior papers have considered directly the numerical scalability of established linear and non-linear methods for solving unbalanced distribution network power flow. This is particularly relevant given the development of new sparse algorithms in the past two decades, some of which have been specifically tailored for circuit analysis \cite{davis2010algorithm}.

The main contribution of this work is to propose an empirical method to estimate complexity coefficient $ \alpha $, then use this approach to tighten the very wide range of $ \alpha $ values reported for $ \Ybus$-based distribution power flow methods. Results indicate `almost linear' complexity--i.e., that $ \alpha $ is superlinear ($ \alpha>1 $), but, much closer to linear than quadratic. Based on this analysis, we argue that it is not unreasonable to say $\alpha \approx 1$.

In this paper, we first describe several $ \Ybus $-based algorithms that can be used for either linear or non-linear power flow (Section~\ref{s:method}). The proposed empirical method is then introduced (Section~\ref{s:regression}) to demonstrate how the value of complexity exponent $ \alpha $ can be estimated. The almost linear scalability of $ \Ybus $-based algorithms is contrasted with worse-than-quadratic complexity of other admittance-like matrices, highlighting how sparsity is, in itself, insufficient to explain the complexity (Section~\ref{s:results}). Conclusions are then drawn (Section~\ref{s:conclusions}).

\textit{Notation.} A backslash operator $ \bs $ solves a (sparse) set of linear equations, with $ \nnz(\cdot) $ counting the number of non-zero elements of a (sparse) matrix. For conciseness of exposition we do not differentiate between power, current and voltage representations of phasors (e.g., polar or rectangular co-ordinates); similarly, we implicitly include or neglect source variables in $ \vbus,\,\ibus $ in definition \eqref{e:ybus_defn}, as can be determined by context.

\section{Distribution Network Power Flow Using the Bus Admittance Matrix}\label{s:method}

The canonical unbalanced distribution network power flow problem is to determine voltage phasors $ v_{\mathrm{bus}} $ based on a given source (or `slack') voltage $ \vsrc $ and known real and reactive power injections at each node $ s_{\mathrm{bus}} $, i.e.,
\begin{equation}\label{e:load_flow_defn}
\vbus = f(\vsrc,\,\sbus)\,.
\end{equation}
Power injections $ \sbus $ are sometimes a function of the voltage at its bus (e.g., for impedance loads). 

The power flow equations $ f $ are non-linear, and so the solution of \eqref{e:load_flow_defn} is determined either using an iterative approach (Section~\ref{ss:nonlinear_scalability}) or via approximation (e.g., a linearization, as considered in Section~\ref{ss:linear_scalability}). To explore how scalability of sparse solve using $ \Ybus $ compares to similar sparse matrices, we then introduce an admittance-like matrix (Section~\ref{ss:impicit_ybus}).

\subsection{Non-Linear Distribution Network Power Flow}\label{ss:nonlinear_scalability}

Iterative Fixed Point-based methods are commonly used by distribution system simulation software to solve \eqref{e:load_flow_defn}, using the iterative rule
\begin{equation}\label{e:nonlinear_update}
\vbus[k+1] = \Ybus \bs i_{\mathrm{cmpstn.}} ( \vbus[k] ) \,,
\end{equation}
where vector $ i_{\mathrm{cmpstn.}} $ is the $ k $th compensation current for all loads \cite{dugan2011open}. The problem \eqref{e:nonlinear_update} iterates until convergence (e.g., when the relative difference between $ \vbus[k+1] $ and $ \vbus[k] $ drops below the pre-specified tolerance).

If the number of iterations required to solve \eqref{e:nonlinear_update} is independent of the scale of the problem, then the solution of non-linear power flow equations will be dependent \textit{only} on the complexity of a sparse solve with the bus admittance matrix $ \Ybus $. In the simulations conducted (Section~\ref{s:results}), this assumption held.

\subsection{Sparse Linear Distribution Network Power Flow}\label{ss:linear_scalability}

As with the non-linear power flow iteration \eqref{e:nonlinear_update}, linear power flow methods often inherit the same sparsity pattern as the bus admittance matrix $ \Ybus $. For example, the fixed-point linear method of \cite{bernstein2018load} can be written in the form
\begin{equation}\label{e:fpl_form}
\vbus = (\Ybus \bs H \sbus ) + \vbus^{0}\,,
\end{equation}
where $ H $ is a sparse matrix (based on the non-linear power flow solution at a chosen linearization point) and $ \vbus^{0} $ is the nominal linearization voltage when $ \sbus=0 $.

The $ \Ybus $ matrix can also dominates the structure of the power flow Jacobian $ D $ (as the change in voltage phasors with respect to power injections). This Jacobian is derived fully in rectangular co-ordinates in \cite{bernstein2018load}. For brevity, the partial derivatives are reproduced fully; it can instead be noted that $ D $ is calculated from a sparse linear system of the form
\begin{equation}\label{e:implicit_DS}
\begin{bmatrix}
S_{11} & S_{12} \\
S_{21} & S_{22} \\
\end{bmatrix}
\begin{bmatrix}
D \\
\tilde{D} \\
\end{bmatrix}
=
\begin{bmatrix}
S_{a} \\
S_{b} \\
\end{bmatrix}\,,
\end{equation}
where $ S_{(\cdot )} $ are matrices of admissible dimension based on partial derivatives of the power flow equations, and $ \tilde{D} $ is the Jacobian of delta-connected load currents with respect to bus powers. An equivalent implicit, sparse linearization that avoids the explicit calculation of $ D $ is therefore
\begin{equation}\label{e:implicit_sln}
\vbus = \begin{bmatrix}
I & 0 \\
\end{bmatrix}
 \left ( \begin{bmatrix}
S_{11} & S_{12} \\
S_{21} & S_{22} \\
\end{bmatrix}
\bs \begin{bmatrix}
S_{a} \\
S_{b} \\
\end{bmatrix}
\sbus \right ) + \vbus^{0}\,,
\end{equation}
where $ I $ and $ 0 $ are identity and zero matrices of admissible dimension.

Note that dense linear forms of both \eqref{e:fpl_form} and \eqref{e:implicit_sln} also exist, requiring only matrix-vector multiplication and addition for linear power flow calculations,
\begin{equation}\label{e:explicit}
\vbus = M\sbus +\vbus^{0}\,,
\end{equation}
where $ M $ is a dense matrix. Dense linear models have time and memory complexity of $ \Oo(n^{2}) $. They therefore scaling poorly as compared to sparse linearizations (e.g., $ M $ requires more than 50~GB memory for $ n = 40,000$).

\subsection{An Admittance-like Sparse Matrix for Comparison}\label{ss:impicit_ybus}

In this subsection, we introduce a sparse random matrix $ \Upsilon (n) \in \mathbb{R}^{n\times n} $ which is superficially similar to the bus admittance matrix $ \Ybus $. This is used in the sequel to show how the time complexity of a sparse solve operation with $ \Ybus $ (e.g., for \eqref{e:nonlinear_update}, \eqref{e:fpl_form} or \eqref{e:implicit_sln}) is much more scalable than solving with other similar sparse matrices. 

Specifically, $ \Upsilon (n) $ is defined as
\begin{equation}\label{e:defn_R}
\Upsilon (n) = \texttt{sprandsym}\left (n,\dfrac{(3p-1)}{n} \right ) + I(n)\Upsilon_{0} \,,
\end{equation}
where $\texttt{sprandsym}(n,d) \in \mathbb{R}^{n\times n}$ is sparse random symmetric matrix with density $ d $ (i.e., $ n^{2}d $ non-zero elements), and uniformly distributed non-zero entries in random positions; $ p\in [1,3] $ represents the number of phases per branch; $ I(n) $ is the $ n\times n $ identity matrix; and, $ \Upsilon_{0} $ is chosen so that $ \Upsilon $ is diagonally dominant. 

The matrix $ \Upsilon $ is similar to the bus admittance matrix $ \Ybus $ in the following senses.
\begin{itemize}
\item The value of $ \nnz(\Upsilon (n)) $ is approximately $3pn $. This can be seen from \eqref{e:defn_R}. A bus admittance matrix $ \Ybus $ in a radial network with $ m $ $ p $-phase buses has $ m-1 $ branches so also has
\begin{align}
\nnz( \Ybus ) &= p^{2}(m + 2(m-1)) \label{e:nnz_Y_full} \\
			&\approx 3pn \label{e:nnz_Y} \,,
\end{align}
where it has been assumed conservatively that the primitive admittance matrices of all branches are dense.
\item The density of both $ \Ybus $ and $ \Upsilon (n) $ is $ \approx 3p/n $.
\item Both matrices are symmetric, diagonally dominant and invertible.
\end{itemize}
For the 75 networks studied in this work (Section~\ref{s:results}), the equivalent value of $ p $ (determined as $ \nnz( \Ybus ) /3n $) was between 1.35 and 3.00, with mean across all networks of 2.09.

\begin{figure}\centering
\subfloat[$ \Ybus $ for IEEE 34 Bus ($ n = 104,\,\nnz=804,\,p = 2.58$) ]{\includegraphics[width=0.22\textwidth]{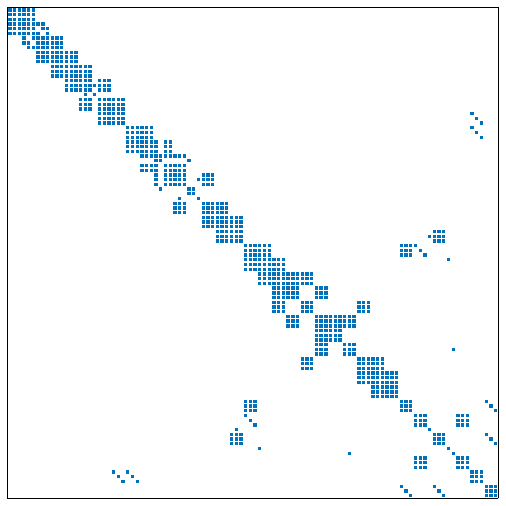}\label{f:spy_ybus}}~
\subfloat[Example $ \Upsilon $ with $ n = 104,\,\nnz=780,\,p = 2.5$.]{\includegraphics[width=0.22\textwidth]{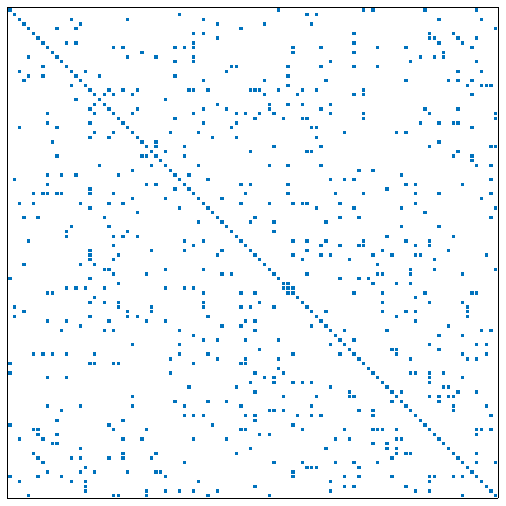}\label{f:spy_ups}}
\caption{Spy plots showing the sparsity pattern of (a) the bus admittance matrix $ \Ybus $ for the IEEE 34 bus network, and (b) an admittance-like sparse matrix $ \Upsilon $ sharing many properties of the bus admittance matrix.}
\label{f:spy_plots}
\end{figure}

An example bus admittance matrix $ \Ybus $ is plotted in Fig.~\ref{f:spy_plots} alongside an admittance-like matrix $ \Upsilon $ with an identical number of nodes $ n $ and a similar number of non-zero elements. It can be seen that $ \Upsilon $ has a very different sparsity pattern to the bus admittance matrix. Nevertheless, the matrices are similar in their properties, as previously noted.

\section{Empirically Estimating Algorithmic Complexity}\label{s:regression}

In this section, we explain how algorithmic complexity is determined empirically via simulations. The general approach is to record the time taken to run a given operation (e.g., solving \eqref{e:implicit_sln}), produce a graph of solution time against the size of a problem, then perform linear regression to determine the complexity coefficient $ \alpha $. If the graph is produced on a log-log axis, the polynomial fit of \eqref{e:complexity_defn} will produce a straight line with gradient of the time complexity coefficient $ \alpha $ as
\begin{equation}\label{e:complexity_logs}
\log (t) = \alpha \log (n) + \mathrm{constant}\,.
\end{equation}
From \eqref{e:complexity_logs}, linear regression is performed with $ \log (n),\, \log(t) $ the independent and dependent variables, respectively.

An advantage of linear regression is that there are standard, well-known methods of estimating quality of fit (through the coefficient of determination, $ r^{2} $) and confidence intervals. Confidence intervals describe a range of values under which the true value of a parameter (in this case the time complexity coefficient) will lie, to a given probability \cite{wasserman2004all}[Ch. 7]. A 95\% confidence interval $ \CI $, assuming a normally distribution of estimates of $ \alpha $ with standard deviation following the standard error $ \sigma $, can therefore be calculated as
\begin{equation}\label{e:conf}
\CI = \alpha \pm 1.96 \times \sigma \,.
\end{equation}
A hypothesis that $ \alpha $ takes a particular value can be rejected if the $ \CI $ does not contain this value--if neither $ \alpha=1 $ or $ \alpha=3 $ are within $ \CI $ then there are grounds to reject a hypotheses that a given algorithm is linear or cubic. Increasing the number of networks that simulations are conducted on will tend to reduce $ \sigma $, resulting in a tighter confidence interval.

Whilst the value of $ n $ is easy to determine, the value to choose for the solution time $ t $ is less clear , as the time taken to run a given algorithm is a random variable. This is because of modern operating systems must allocate computing resource on-the-fly, causing variable delays in performing computational tasks. For the purposes of this work, the solution time $ t $ is simply considered as the median of ten runs of a given algorithm (similarly to \cite{deakin2021analysis}). All operations use a PC with Intel Core i7-8665U and 16~GB memory.

\section{Results}\label{s:results}

In this section, we use the admittance matrix-based power flows methods (Section~\ref{s:method}) and regression (Section~\ref{s:regression}) to estimate the time complexity $ \alpha $ of the core sparse solve operations these methods require. To achieve this over a wide range of $ n $, a library of 75 networks have been collected from five sources (Table~\ref{t:tblNetworksSummary}). Between them, the size of these network covers nearly four orders of magnitude, and includes both North American and European-style networks.

\begin{table}
\centering
\caption{Network libraries used (a total of 75 networks simulated).}\label{t:tblNetworksSummary}
\begin{tabular}{lllll}
\toprule
\multirow{2}*{\vspace{-0.5em} Library [Ref.] } & $ \multirow{2}*{\vspace{-0.5em} Networks (count)} $ & \multicolumn{3}{c}{Nodes $n$}\\
  \cmidrule(l{0.6em}r{0.9em}){3-5}
    &  & Min. & Med. & Max. \\
    \midrule
\begin{tabular}[x]{@{}l@{}}OpenDSS\\library \cite{dugan2011open}\end{tabular} & \begin{tabular}[x]{@{}l@{}}IEEE test feeders;\\EPRI J1, K1, M1;\\EPRI 5, 7, 24 (12)\end{tabular} & 50 & 2589 & 8543 \\
LVNS \cite{enwl2015low} & \begin{tabular}[x]{@{}l@{}}Networks 1-25,\\combined fdrs. (25)\end{tabular} & 4974 & 13389 & 55536 \\
\begin{tabular}[x]{@{}l@{}}European\\MV-LV \cite{deakin2021hybrid}\end{tabular} & \begin{tabular}[x]{@{}l@{}}UG, UG-OHa (2)\end{tabular} & 86448 & 99667 & 112887 \\
\begin{tabular}[x]{@{}l@{}}US synthetic\\networks \cite{mateo2020building}\end{tabular} & \begin{tabular}[x]{@{}l@{}}GSO; SFO P1U, P1R,\\P2U, P2R (7)\end{tabular} & 22051 & 151818 & 314912 \\
\begin{tabular}[x]{@{}l@{}}Non-synthetic\\EULV \cite{deakin2023partitioned}\end{tabular} & \begin{tabular}[x]{@{}l@{}}Networks 1-28\\and Network 30 (29)\end{tabular} & 393 & 1062 & 2091 \\
\bottomrule
\end{tabular}

\label{t:tblNetworksSummary}
\end{table}

\subsection{Case Studies}\label{ss:results_case_studies}

To explore scalability of admittance matrix-based distribution power flow, we explore the behaviour of three methods.
\begin{enumerate}[label=(\roman*)]
\item Non-linear power flow complexity is explored using the default fixed-point algorithm of OpenDSS (which uses the sparse linear solver KLU \cite{davis2010algorithm}). Both the total time to solve and number of iterations are recorded; the time $ t $ is calculated on a per-iteration basis, based on a step change in all load from 60\% to 30\%.
\item Linear power flow is calculated in OpenDSS based on a constant admittance load (and so the solution is found as one sparse solve with the bus admittance matrix).
\item Linear power flow using the implicit (sparse) power flow Jacobian \eqref{e:implicit_DS} (the full system of equations are reported in \cite{bernstein2018load}). This is solved using the MATLAB `backslash' operator \texttt{mldivide}.
\end{enumerate}
In addition, the scalability of two further closely related problems are considered using \texttt{mldivide}:
\begin{enumerate}[label=(\roman*)]
\setcounter{enumi}{3}
\item Sparse linear solve for the bus admittance matrix $ \Ybus $,
\item Sparse linear solve for the sparse admittance-like matrix $ \Upsilon(n) $, explored for $ p \in \{1,\,2,\,3\}$. When solving with $ \Upsilon(n) $, a new matrix is drawn from \eqref{e:defn_R} for each sparse solve (not included in the time to solve $ t $).
\end{enumerate}

\subsubsection{Complexity of Linear and Non-linear Load Flow}
\begin{figure}\centering
\subfloat[Method (iii);  \texttt{mldivide}]{\includegraphics[width=0.192\textwidth]{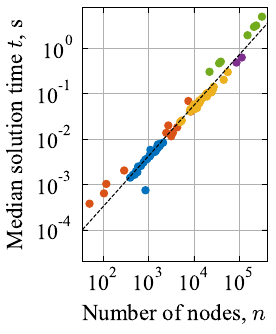}\label{f:pltOpenDSSscalability_fot_matlab}} 
~
\subfloat[Method (i); OpenDSS/KLU]{\includegraphics[width=0.288\textwidth]{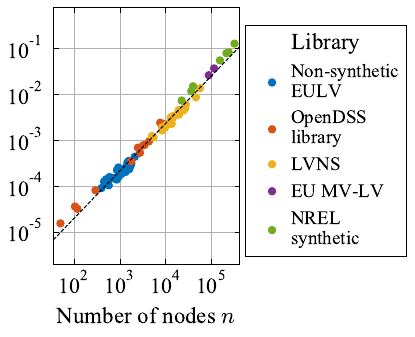}\label{f:pltOpenDSSscalability}} 
\caption{The number of nodes $ n $ against the median of ten solution times $ t $, using (a) the implicit sparse Jacobian \eqref{e:implicit_sln} using \texttt{mldivide}, and (b) the time per iteration for the load flow solution calculated using OpenDSS \eqref{e:load_flow_defn}.}
\label{f:pltScalability}
\end{figure}

Fig.~\ref{f:pltScalability} plots the number of nodes $ n $ against the median time to solve $ t $ for both a linear (Fig.~\ref{f:pltOpenDSSscalability_fot_matlab}, method (iii)) and non-linear method (Fig.~\ref{f:pltOpenDSSscalability}, method (i)), with the dashed line from the estimated linear fit \eqref{e:complexity_logs}. It can be observed that in both cases that the complexity coefficient $ \alpha $ is almost linear--increasing the number of nodes $ n $ from $ 10^{2} $ to $ 10^{5} $ increases time $ t $ by a factor close to $ 10^{3} $. Note that OpenDSS's solve (via KLU) is around ten times faster per iteration than \texttt{mldivide}. This is assumed to be because KLU is designed for circuit problems, where \texttt{mldivide} is a family of general purpose sparse solvers.

Note the presence of outliers, e.g., one Non-synthetic EULV network has a solution time $ t $ which is faster than the expected value based on the fit (in Fig.~\ref{f:pltOpenDSSscalability_fot_matlab}). We also report a small numbers of outliers have also been observed when using other algorithms (not shown). Such variation is seen in other sparse circuit problems \cite{davis2010algorithm}, and so is not unexpected. Future work could explore why some networks might have faster or slower solve times than the general trend.

\begin{figure}
\centering
\includegraphics[width=0.4\textwidth]{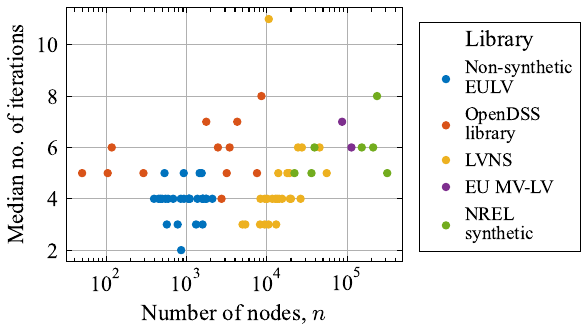}
\caption{The median number of iterations to solve non-linear power flow using OpenDSS is not closely linked to the number of nodes $ n $.}
\label{f:pltNiter}
\end{figure}

The number of iterations required for non-linear power flow convergence (method (i)) is shown in Fig.~\ref{f:pltNiter}. The number of iterations does not substantially increase with number of nodes $ n $. The Non-synthetic EULV library are generally solved in a smaller number of iterations than the other networks, potentially due to lower loading than other small networks which have been designed to be difficult to solve (e.g., IEEE test cases). Taken together, the per-iteration median time to solve and median number of iterations together point to almost-linear complexity of non-linear power flow of OpenDSS, in sharp contrast to the cubic complexity reported in \cite{carreno2022log}.

\subsubsection{Comparing complexity of algorithms (i)--(v)}
The estimated complexity coefficient $ \alpha $ of all five algorithms (i)-(v) are presented in Table~\ref{t:tblResults}. Results for algorithms (ii) and (iv) show qualitatively a similar fit to the results to those of (i) and (iii) (Fig.~\ref{f:pltScalability}) and so for conciseness are not plotted. From this table, it can be concluded that the hypothesis that the complexity is cubic can be rejected. However, it is also interesting to note that the hypothesis that the scalability is linear can \emph{also} be rejected. Instead, the four algorithms show complexity that is close to linear, and it is suggested these algorithms might be referred to has having `almost linear' complexity (not to be confused with `nearly linear' complexity, $ \mathcal{O}\left ( n\log (n) \right ) $).

\begin{table}
\centering
\caption{Linear regression \eqref{e:complexity_logs} results for five sparse problems.}\label{t:tblResults}
\begin{tabular}{lllll}
\toprule
ID & Description & \tcell{Complexity\\coeff., $\alpha $} & \tcell{Standard \\error, $ \sigma $} & \tcell{Coeff. of\\det., $r^{2} $} \\
\midrule
(i) & \tcell{OpenDSS, non-linear\\(per iteration)} & 1.037 & 0.014 & 0.987 \\
(ii) & \tcell{OpenDSS, constant\\admittance load} & 1.068 & 0.011 & 0.993 \\
(iii) & \tcell{Jacobian (sparse\\linear solve)} & 1.116 & 0.023 & 0.971 \\
(iv) & \tcell{Admittance matrix\\(sparse linear solve)} & 1.091 & 0.014 & 0.988 \\
(v) & \tcell{Admittance-like $\Upsilon(n)$\\(sparse linear solve)} & $\geq$ 2 & - & - \\
\bottomrule
\end{tabular}

\label{t:tblResults}
\end{table}

\begin{figure}
\centering
\includegraphics[width=0.31\textwidth]{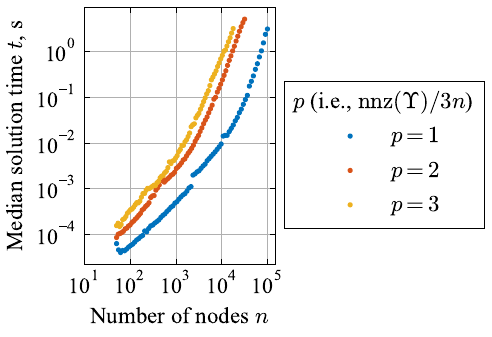}
\caption{The number of nodes $ n $ against the median solution time $ t $ for algorithm (v) using the admittance-like sparse matrices $ \Upsilon(n) $.}
\label{f:check_sprand_complexity}
\end{figure}

A graph of the time to solve against number of nodes for algorithm (v) is plotted in Fig.~\ref{f:check_sprand_complexity}. Unlike algorithms (i)-(iv), this shows qualitatively a poor fit against the polynomial complexity model \eqref{e:complexity_logs}, as the gradient increases noticeably with $ n $. Therefore, polynomial complexity coefficients (as in \eqref{e:complexity_defn}) are only valid locally. The complexity coefficient $ \alpha $ was greater than 2.46 for all $ p\in\{1,\,2,\,3\} $ with coefficient of determination $ r^{2} \geq 0.977$; hence, the overall complexity $ \alpha $ is therefore reported as greater than quadratic in Table~\ref{t:tblResults}. These coefficients were calculated considering 11 logarithmically spaced values of $ n $ between 3,000 and 30,000 (for $ p=2,3 $) and between 10,000 and 100,000 (for $ p=1 $). In summary, whilst sparsity clearly is linked to the scalability of admittance matrix-based power flow, it is not sufficient to explain the almost linear complexity of algorithms (i)-(iv). The fact that circuit matrices are better-suited to be solved than other sparse matrices with a similar density has been noted in \cite{davis2010algorithm}.

\subsection{Discussion}\label{ss:results_discussion}

Two main future directions are suggested. Firstly, the empirical approach considered in this work using \eqref{e:complexity_logs} could be complemented by theoretical analysis to explore in more detail how modern sparse direct or iterative linear solvers scale with a variety of power flow solution methods. For example, direct sparse solution methods for regular finite element mesh grids have a known complexity \cite{george1973nested}--future work could explore if a similar bound exists for sparse solve operations using $ \Ybus $.

Secondly, the fact of almost-linear complexity of standard power flow algorithms is needed to properly contextualize the potential benefits of power flow acceleration approaches for solving the basic load flow problem \eqref{e:load_flow_defn} or other derivative load flow-based problems (e.g., time series analysis or optimal power flow (OPF) problems). For example, there is potential to exploit multi-threading in modern computing architectures alongside sparse solvers, using network decompositions (or `tearing') for parallelization \cite{montenegro2022simplified}. Other approaches have explored block-sparse formulations for fast matrix-vector calculations as part of probabilistic load flow \cite{deakin2021analysis}, or improved scalability and acceleration of OPF problems via permutation to bordered block-diagonal form \cite{kardovs2022beltistos}. 

Finally, we note that using a library of networks (on the scale of those reported in Table~\ref{t:tblNetworksSummary}) can improve robustness of estimates of scalability. The 75 network models considered in this work more closely mirrors the scale of efforts seen in benchmarking in other computational fields (e.g., 81 circuit matrices used to explore algorithmic performance in \cite{davis2010algorithm}).

\section{Conclusion}\label{s:conclusions}

Distribution network analysis based on the bus admittance matrix has been shown empirically to be very scalable, having almost linear computational complexity for both linear and non-linear solution methods. There are many practical and theoretical aspects of distribution power flow that can be the topic of future research for large-scale networks. Simulations using a small number of networks, as has conventionally been considered, only hint at an algorithm's complexity. In contrast, the proposed empirical approach assesses the time complexity coefficient systematically. It is concluded that such algorithmic benchmarking will be crucial for comparing promising power flow analysis methods for tackling the large-scale network analysis tasks increasingly required by network operators.


\end{document}